\documentstyle[aps]{revtex}

\draft

\begin{document}

\title{Correlated electrons and generalized statistics}

\author{Qiuping A. Wang}
\address{Institut Sup\'erieur des Mat\'eriaux du Mans,\\ 44, Avenue F.A.
Bartholdi, 72000 Le Mans, France}


\maketitle

\begin{abstract}
Several important generalizations of Fermi-Dirac distribution are compared to numerical and experimental
results for correlated electron systems. It is found that the quantum distributions based on incomplete
information hypothesis can be useful for describing this kind of systems. We show that the additive incomplete
fermion distribution gives very good description of weakly correlated electrons and that the nonadditive one is
suitable to very strong correlated cases.
\end{abstract}



\section{Introduction}

In this work, we attempt to relate generalized statistical
theories to correlated electron or quasiparticle systems which,
since 1980's, have been subjected to intensive theoretical and
experimental studies due to their important roles in
optical\cite{Hors01}, mechanical\cite{Rout99}, electric and
magnetic \cite{Rice94,Bock99,Corelec1,Corelec2,Corelec2a,Corelec3}
properties of metal compounds with partially filled d- or
f-electrons. In general, with complex correlations, the
conventional Boltzmann-Gibbs statistics (BGS) for free fermions is
no longer valid. Most of the theoretical study up to now was
numerical simulations based on approximate models including
essential physics of the systems of interest. Although these
theoretical works are very important and fruitful, the need of
general understanding and theoretical formulation of the
distribution laws begins to be felt. Due to the complexity of the
correlations, we can conjecture that generalization of Fermi-Dirac
(FD) theory may be useful. In this sense, some efforts have been
made to understand localized electron behavior on the basis of
fractal geometry (see \cite{Levy87} and references there-in), a
generalization of normal space-time.

In what follows, I present a work in this direction based, not on
the fractal or chaotic assumption, but on generalized statistical
theories proposed for complex systems showing correlated
phenomena. The theoretical results will be compared to numerical
and experimental ones for correlated electrons. It is expected
that this attempt can help to better understand different
generalized theories and identify valid theories for this special
circumstance.

\section{Some important generalizations of BGS and correlated electrons}

Indeed, in the last decade, BGS theory has experienced a turbulent
period with the rapid development of some anomalous theories to
treat complex systems. These theories are anomalous in the sense
that they may perturb our old conceptions concerning, e.g.,
information, energy, quantum states, additivity etc. The main
characters of these theories can be resumed as follows :

1) Consideration of nonadditivity in entropy, energy, quantum
occupation number etc. depending on empirical parameters, e.g.
$0<q<\infty$ in Tsallis nonextensive statistical mechanics
(NSM)\cite{Tsal88} and in its incomplete statistics (IS)
version\cite{Wang00,Wang00a,Wang01,Wang02,Wang02a} as well as in
quantum group theory (QGT)\cite{Gruv99,Mont93}, $\kappa\geq 0$ in
$\kappa$-statistics (KS)\cite{Kani01}.

2) Bosonization of fermions or fermionization of bosons allowing
intermediate occupation number related to empirical parameter,
e.g. $0\leq\alpha\leq 1$ in fractional exclusion
statistics(FES)\cite{Mont93,Wu01,Hald91,Kani96,Buyu01} and
$-1\leq\eta\leq 1$ for the FES in the frame of KS\cite{Kani01}.

3) The empirical parameters take some particular values when the theories
recover BGS ($q=1$, $\alpha=0,1$, $\kappa=0$ and $\eta=-1, 0, 1$).

The philosophy of these generalizations, explicit or not, is to
introduce empirical parameters to ``absorb" the effects of complex
correlations. So interacting systems can be mathematically treated
as noninteracting or conventional ones. For example, the total
energy of an interacting ``free particle" can be written as
$p^2/2m$ where $p$ is the momentum and $m$ the mass of the
particle. The extra interaction energy is in this way ``absorbed"
in the empirical parameter when they are different from the values
corresponding to conventional cases.

About the generalization of FD statistics, Figure 1 shows the
different fermion distributions at $T$=100 K of the generalized
statistical mechanics mentioned above. We see that the
discontinuity in occupation number $n$ at Fermi energy $e_f$
depends almost only on temperature for all these distributions.
The sharp $n$-drop hardly changes for whatever value of the
empirical parameters. In addition, we note that the approximate
distribution (AD) $n=\frac{1}{[1+(q-1)\beta(e-e_f)]^{1/(q-1)}+1}$
($0<q\leq 1$ in order that the inverse temperature $\beta$ can go
to $\infty$) of NSM obtained with factorization
approximation\cite{Buyu01,Buyu93} by neglecting correlation energy
between the particles is only slightly different from FD one. So
its Fermi energy is almost identical to that of FD. The situation
of KS is similar. At low temperature, the $\kappa$-distribution of
standard fermion ($\eta=1$)
$n=\frac{1}{[\sqrt{1+\kappa^2\beta^2(e-e_f)^2}
+\kappa\beta(e-e_f)]^{1/\kappa}+1}$\cite{Kani01} is hardly
different from FD for $\kappa$ values which give finite internal
energy (e.g. $0\leq\kappa<1/2$ for two dimensional fermion gas).
For FES fermion distribution
$(1/n-\alpha)^\alpha(1/n-\alpha+1)^{1-\alpha}=e^{(e-e_f)/kT}$\cite{Wu01}
(or equivalently $n=\frac{1}{e^{q\beta(e-e_f)}-1}-
\frac{(1+\alpha)/\alpha}
{e^{q\beta(e-e_f)(1+\alpha)/\alpha}-1}$\cite{Wang02b}), the
$n$-discontinuity depends only on temperature just as for FD. The
influence of $\alpha$ smaller than unity (for FD) is to decrease
$e_f$ ($\alpha$ times) and to increase the maximal occupation
number ($1/\alpha$). As for the nonextensive fermion distribution
of IS
$n=\frac{1}{[1+(q-1)\beta(e-e_f)]^{q/(q-1)}+1}$\cite{Wang02,Wang02a},
though $n$ drop at $e_f$ is always very sharp, $e_f$ increase
considerably as $q$ decreases from unity. And consequently, there
is an important wide $n$ decrease with decreasing $q$. It was
shown\cite{Wang02a} that, for $q\rightarrow 0$, $e_f$ can increase
up to two times the $e_f$ of FD with $n=1/2$ ($n=0$) for all
energy below (above) $e_f$. Another common character of NSM and IS
distributions is that, due to the energy cutoff with $q<1$, there
are few electrons above $e_f$ at low temperature (large $\beta$).

Now let see the distributions of correlated electrons. From some results of experiments and numerical
simulations based on low dimension Kondo lattice models (KLM)\cite{Corelec1,Corelec2,Corelec2a,Corelec3}, we
notice two important effects of correlations. First, the wide decrease of $n$ and the sharp $n$ drop at $e_f$
much larger than that without correlation have indeed been observed for very strongly correlated electrons
(with the coupling parameter $J\geq 4$)\cite{Corelec2,Corelec3} (see Figure 3 and 4). So it is possible to
describe correlated electron systems in strong coupling regime by IS fermion distribution with $q$ very smaller
than unity (close to zero)\cite{Wang02a}. On the other hand, another effect of correlation, in the weak
coupling regime, is the flattening of $n$ drop at $e_f$\cite{Corelec1,Corelec2,Corelec2a,Corelec3}. That is
that the correlation, even at low temperature, drives electrons above $e_f$ so that the $n$ discontinuity
becomes less and less sharp as the correlation increases. Curiously, this flattening of $n$ discontinuity at
$e_f$, though confirmed by experimental results\cite{Corelec1}, is completely absent in the fermion
distributions given by all the statistics, extensive (FES) or nonextensive (NSM, IS and KS), mentioned above.

Fermi surface plays an essential role in the physical properties of metals and metal compounds. So it is of
great interest for generalized statistical theories to take into account the essential physics relevant to Fermi
energy. In what follows, we will show that it is possible to cover the observed weak correlation effects upon
$e_f$ within an additive generalization of BGS based on the IS principles. The reader can find a complete review
of IS theory in references \cite{Wang00,Wang00a,Wang01,Wang02,Wang02a}.

\section{Additive incomplete statistics}

The concept of IS is inspired by incomplete probability distribution (i.e. $\texttt{Tr}\rho=Q\neq
1$)\cite{Reny66} as well as the theoretical difficulties one encounters with $\texttt{Tr}\rho^q\neq 1$ in
NSM\cite{Tsal88,Wang01}. The basic assumption of IS is that our knowledge about physical systems is in general
incomplete due to unknown interactions or their effects which can not be studied explicitly. In this case, there
are always missing informations about the physical systems. So probability distribution is in general inexact
and incomplete and can not sum to one. One should write $\texttt{Tr}(\rho/Q)=\texttt{Tr}F(\rho)=1$ where $F$ is
certain function to be determined. In the case of complete or approximately complete distribution (such as BGS),
$F$ is identity function. In my previous paper, in order to keep Tsallis framework of NSM, I proposed
$F(\rho)=\rho^q$ so that
\begin{equation}                        \label{2}
\texttt{Tr}\rho^q=1,
\end{equation}
where $q$ is Tsallis entropy index\cite{Wang00,Wang00a,Wang01}. Since $\rho<1$, we have to set $q\in[0,\infty]$.
When $q=1$, $\rho$ becomes complete, which implies a complete or quasi-complete knowledge about the system of
interest. The nonextensive IS formalism\cite{Wang00,Wang01} on the basis of Tsallis entropy yields Tsallis
$q$-exponential distributions and the fermion distribution\cite{Wang02,Wang02a} discussed above. An successful
application of Eq.(\ref{2}) to the study of some power laws on the basis of R\'enyi entropy\cite{Reny66} can be
found in reference \cite{Bash00}. One can even find a plausible justification of Eq.(\ref{2}) in a work of
Tsallis on the probability distribution on some simple fractal supports like Koch or Cantor sets\cite{Tsal95}.
Indeed, the phase space of the systems including complex interactions is in general chaotic or fractal
structure. In these structures, exact probability distribution is impossible due to the fact that, first, the
structures are not differentiable and integrable, second, as mentioned above, some state points in the
structures are not accessible. So the treatment of the incomplete distributions, not in the fractal space of
$d_f$ dimension, but in a differentiable and integrable $d$-dimension space, is inevitable. This consideration
yields in a natural way the incomplete normalization Eq.(\ref{2})\cite{Wang02a}.

Now we suppose that the system of interest has additive information and physical quantities (e.g. $H=\sum_iH_i$
where $H$ is the hamiltonian of the compound system and $H_i$ that of the $i^{th}$ subsystem). This case may
happen if there are only weak and short range interactions, so that the Hartley formula $\ln(1/\rho)$ for
information measure should apply. The entropy of the system can be defined \`a la Shannon as
follows\cite{Wang00a} :
\begin{equation}                        \label{3}
S=k\texttt{Tr}\rho^q\ln(1/\rho).
\end{equation}
$S$ obviously becomes Gibbs-Shannon entropy ($S_{GS}$) when $q=1$, which
identifies $k$ to Boltzmann constant. It is straightforward to see that $S$
verifies all the properties of $S_{GS}$. For {\it microcanonical ensemble}
($\rho^q=\frac{1}{w}$), we have $S=\frac{k}{q}lnw$ which decreases with
increasing $q$ value. In general, $\Delta S=S-S_{GS}<0$ (or $>0$) if $q>1$ (or
$q<0$).

For {\it canonical ensemble}, we postulate
\begin{equation}                    \label{4}
U=\texttt{Tr}\rho^qH
\end{equation}
where $H$ is the Hamiltonian and $U$ the internal energy of the system. The maximum entropy of $S$ with
Eq.(\ref{2}) and (\ref{4}) as constraints leads to :
\begin{equation}                        \label{5}
\rho=\frac{1}{Z}e^{-\beta H}
\end{equation}
with $Z=(\texttt{Tr}e^{-q\beta H})^{1/q}$. The Lagrange parameter $\beta$ can be determined by $\frac{\partial
S_q}{\partial U_q}=k\beta=\frac{1}{T}$ where $T$ is the absolute temperature. It is easy to see that all
thermodynamic relations in this theory are identical to those in BGS.

We indicate in passing that this generalization of BGS is obviously the consequence of the
``power-normalization" Eq.(\ref{2}). Other useful normalizations, if any, of incomplete distributions may lead
to different generalization of BGS. For example, if we use Hartley formula as information measure and define
expectation value for additive entropy and energy with $F(p_i)$ satisfying $\frac{\partial \ln F(x)}{\partial
x}=\frac{1/x}{\alpha-\ln x-(-\ln Zx)^{1/\gamma}}$ where $\gamma$ is an empirical parameter, maximum entropy will
lead to the famous {\it stretched exponential distribution}\cite{Jund01} $p_i=\frac{1}{Z}e^{-(\beta
E_i)^\gamma}$ where $E_i$ is positive energy of the system at state $i$, $\alpha$ and $\beta$ the Lagrange
multipliers related respectively to normalization and energy constraint $U=\sum_{i=1}^{w}F(p_i)E_i$. When
$\gamma=1$, we can recover $\frac{\partial \ln F(x)}{\partial x}=\frac{1}{x}$ as in BGS.

\section{Additive IS Fermion distribution}

For {\it grand canonical ensemble}, the same machinery as above leads to
\begin{equation}                    \label{9}
\rho=\frac{1}{Z}e^{-\beta(H-\mu N)}
\end{equation}
with $Z=[\texttt{Tr} e^{-q\beta(H-\mu N)}]^{1/q}$. Supposing $U=\sum_jn_je_j$ and $N=\sum_jn_j$ in accordance
with the additivity assumption, the average occupation number $n$ of one-particle state of energy $e$ can be
straightforwardly calculated\cite{Wang00a} :

\begin{equation}                                 \label{12}
n=\frac{1}{e^{q(e-e_f)/kT}+1}
\end{equation}
which recovers FD distribution for $q=1$. The Fermi energy can be calculated in the standard way. For 2-D
systems, e.g., we have $e_f=e_{f_0}+k_qT\ln(1-e^{-e_f/k_qT})$. In Figure 2 is plotted the $T$-dependence of
$e_f$ for different $q$ values for 1-D fermion systems. We see that $e_f$ considerably changes with decreasing
$q$. From Eq.(\ref{12}), we see that $q$-dependence of $e_f$ at given $T$ will be similar to the $T$-dependence.
The effect of $q$ on other physical quantities (specific heat, electrical conductivity, susceptibility,
effective mass etc.) is discussed in \cite{Wang00a}.

The $n$ distribution around the Fermi momentum $k_f$ can be estimated from Eq.(\ref{12}) using
$e=\hbar^2k^2/2m$. We get :
\begin{equation}                                \label{14}
n=\frac{1}{2}[1+\frac{q}{kT} \frac{\hbar^2k_f^2}{m}(k-k_f)]
\end{equation}
which is similar to the analog given by a Monte-Carlo calculation based on a
tight-binding Kondo lattice ($KL$)\cite{Corelec2a}. A comparison with \cite{Corelec2a}
leads to $Z=\frac{q\hbar k_f}{2}$, where $Z$ is the quasiparticle weight of
photoemission spectrum. Note that $Z$ is $k_f$-dependent here. Eq.(\ref{14}) tells us
that decreasing $q$ yields a flattening of $n$ drop at $e_f$.

In Figure 3 and 4, we compare the momentum distribution given by Eq.(\ref{12}) for $T$=50 K to some numerical
results of the momentum distribution of 1-D correlated electrons\cite{Corelec2,Corelec3}. We see that, for
about $J<1$ and $q>0.003$, KLM calculations can be well reproduced by Eq.(\ref{12}). But for smaller $q$ or
stronger coupling with $J>1$, a long tail in the KLM distributions begins to develop at high energy and can not
be reproduced by Eq.(\ref{12}). At the same time, a new Fermi surface at $k_f+\pi/2$ starts to appear and a
sharp $n$ drop (energy cutoff) takes place at the new Fermi momentum. All these strong correlation effects turn
out to be completely absent within the present IS Fermion distribution. It is clear that this discrepancy marks
the limit of this formalism and perhaps implies that nonextensivity is no more negligible in very
strong-coupling case. It is worth noticing that the strong correlation effects, absent in additive IS fermion
distribution, are just what we noted in nonextensive IS fermion distribution\cite{Wang02a} plotted in Figure 1.
This suggests that a combination of these two IS generalizations of BGS, representing respectively two
different aspect of correlation, may be an interesting approach leading to more generally valid statistical
mechanics suitable for correlated particle systems.

\section{conclusion}
We have compared several important generalizations of FD distribution to numerical and experimental results for
correlated electron systems. It is found that the generalized fermion distributions based on incomplete
information can be useful for describing these systems. It is shown that the extensive IS fermion distribution
gives a good description of correlated electrons in weak coupling regime. On the other hand, it fails to
describe strongly correlated conduction electrons and localized $f$-electrons which are possible to be
described by the nonextensive IS fermion distribution. The latter, though powerless for the description of
weakly correlated electrons, shows similar behavior as the heavy electrons in strong coupling regime : strong
increase of $e_f$ and $n$-cutoff accompanied by a general decrease in $n$ at all energy up to $e_f$. In a
current work, we are trying to combine these two partially valid generalizations. Further results will be
presented in a future paper.

{\Large Figure caption :}

\begin{figure}[p] \label{f1}

\caption{Fermion distributions of FD, FES, AD, KS and nonextensive IS at 100 K.
The fermion density is chosen to give $e_f^0 = 1$ eV for FD distribution at
$T=0$. We note that all these distributions show sharp $n$ drop at Fermi energy
$e_f$. In addition, AD and KS distributions are only slightly different from
the FD one even with $q$ very different from unity and maximal $\kappa$ far
from zero. On the other hand, the nonextensive IS distribution changes
drastically with decreasing $q$. We notice a wide decrease of $n$ and a strong
increase of $e_f$ with decreasing $q$. As $q\rightarrow 0$, the occupation
number tends to 1/2 (zero) for all states below (above) $e_f$ which increases
up to 2 times $e_{f_0}$, a fact shown by numerical calculations for strong
correlated electrons.}
\end{figure}

\begin{figure}[p] \label{f2}

\caption{$T$-dependence of Fermi momentum $k_f$ of the extensive IS fermion distribution in present work. The
density of fermions is chosen to give $k_{f_0}=0.25\pi$ in the first Brillouin zone. The $T$-dependence of $k_f$
is in general not monotonic, in contrast with the classical decreasing behavior of $e_f$ with increasing
temperature. We notice that, at low temperature, $k_f$ shows an increase with increasing $T$.}
\end{figure}

\begin{figure}[p] \label{f3}

\caption{Comparison of extensive IS fermion distribution (lines) with the numerical results (symbols) of Eder el
al on the basis of  Kondo lattice $t-J$ model (KLM) for different coupling constant $J$ [6]. The density of
electrons is chosen to give $k_{f_0}=0.25\pi$ in the first Brillouin zone. We note that IS distribution
reproduces well the numerical results for about $J<1$. When coupling is stronger, a long tail in the KLM
distributions begins to develop at high energy.  At the same time, a new Fermi surface at
$k=k_{f_0}+\pi/2=0.75\pi$ starts to appear and a sharp $n$ drop takes place at the new Fermi momentum. At $J=4$,
KLM distribution (x-marks) is very different from that of IS (e.g. $q=0.0011$). The solid line fitting better
the $J=4$ KLM distribution is given by the IS version of fractional exclusion distribution
$(1/n-\alpha)^\alpha(1/n-\alpha+1)^{1-\alpha}=e^{q\beta(e-e_f)}$ [19,20] or $n=\frac{1}{e^{q\beta(e-e_f)}-1}-
\frac{(1+\alpha)/\alpha} {e^{q\beta(e-e_f)(1+\alpha)/\alpha}-1}$ [24] with $1/\alpha=0.85$ due to the KLM
occupation number smaller than 0.5 at low momentum $k$.}
\end{figure}

\begin{figure}[p] \label{f4}

\caption{Comparison of extensive IS fermion distribution (lines) with the numerical results (symbols) of
Moukouri el al on the basis of Kondo lattice $t-J$ model (KLM) for different coupling constant $J$ [8]. The
density of electrons is chosen to give $k_{f_0}=0.35\pi$ in the first Brillouin zone. We note that IS
distribution reproduces well the numerical results for about $J<2$. When the coupling is stronger, a long tail
in the KLM distributions begins to develop at high energy. A new Fermi surface with sharp $n$ drop starts to
appear at about $k=0.7\pi$, an energy cutoff effect absent in the extensive IS fermion distribution.}
\end{figure}

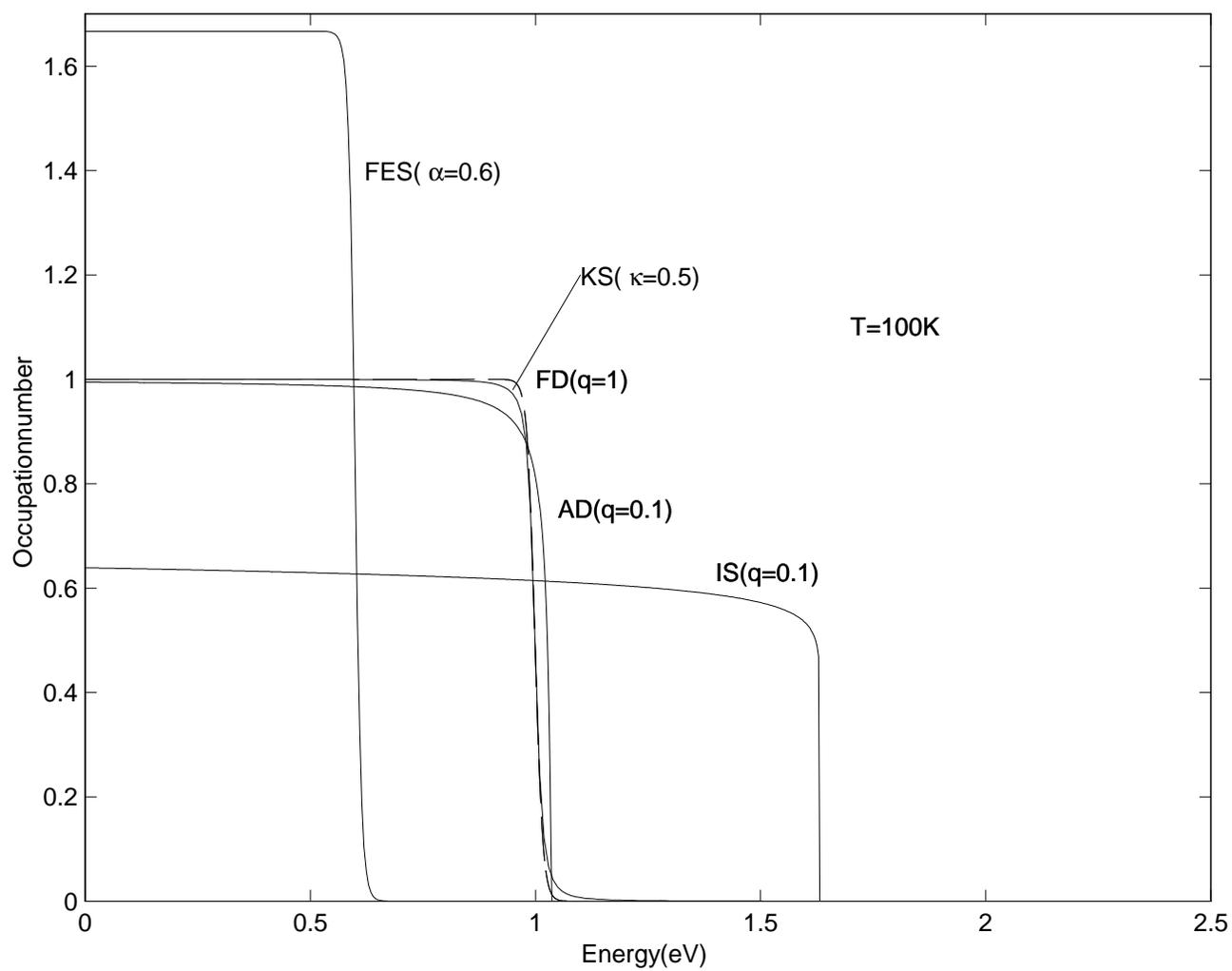
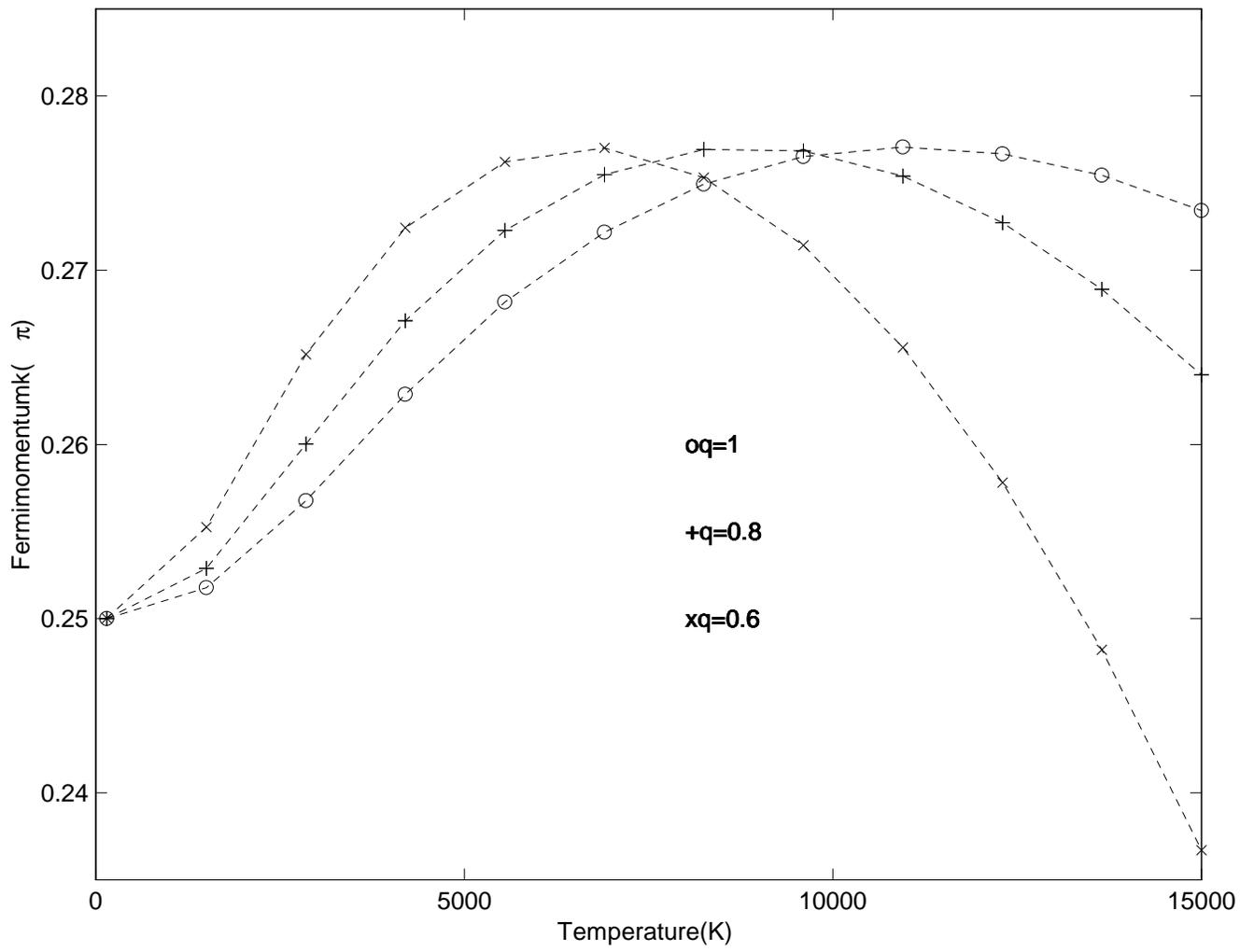
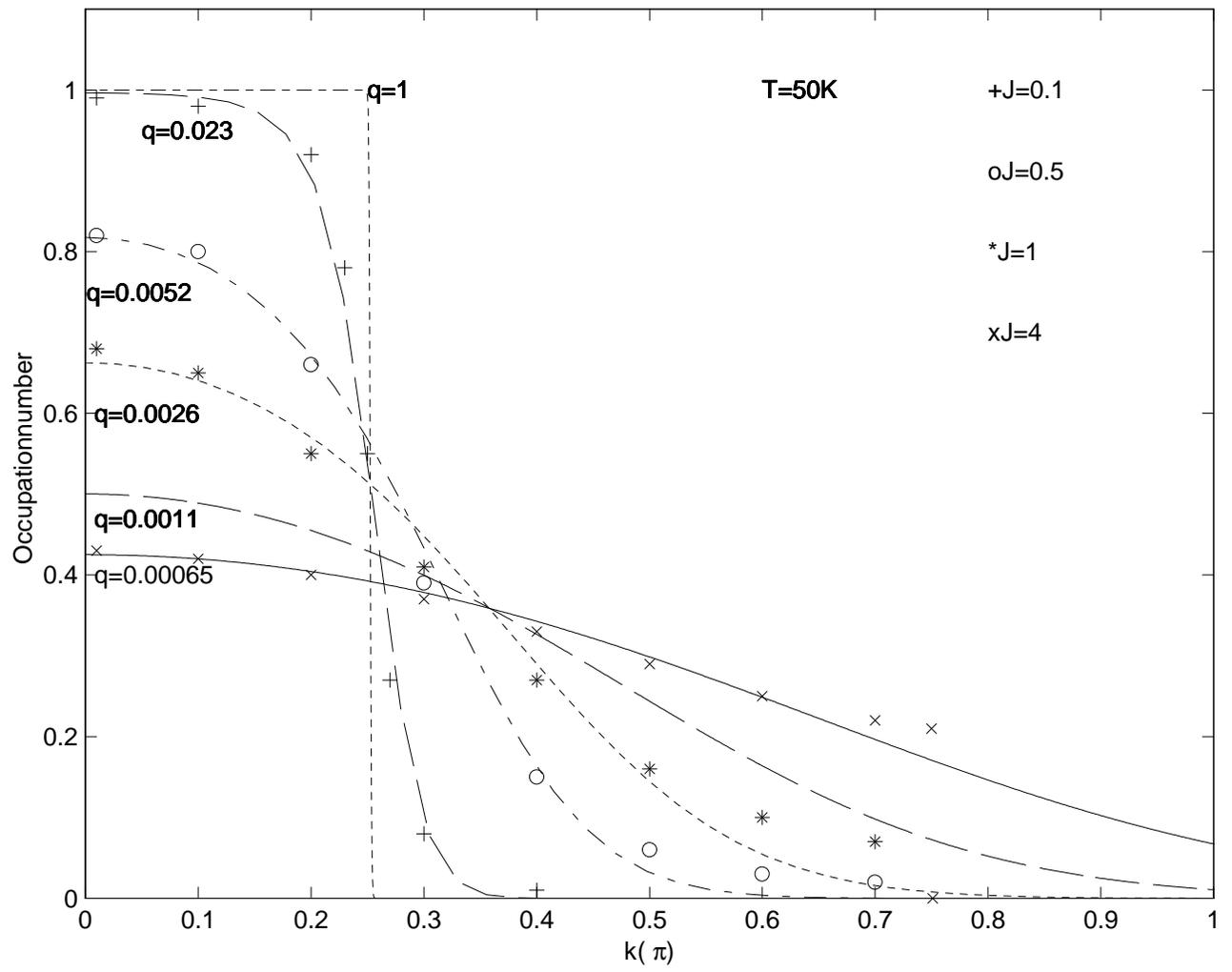
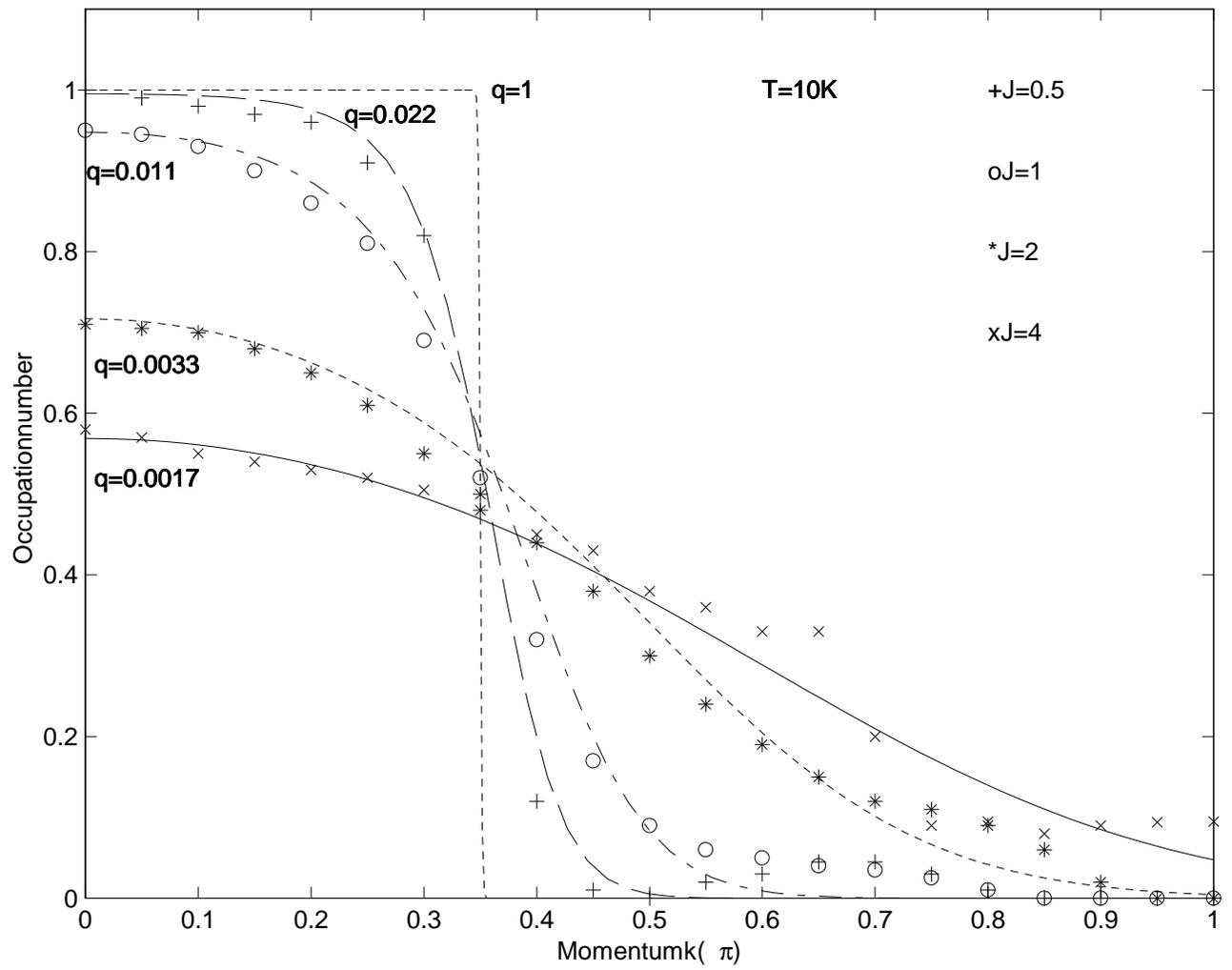
\end{document}